\documentclass[10pt,conference]{IEEEtran}

\usepackage{tikz}
\usetikzlibrary{arrows.meta,backgrounds,calc,fit,positioning}

\definecolor{figaccent}{HTML}{003661}
\definecolor{figink}{HTML}{0B0B0B}
\definecolor{figmuted}{HTML}{4F5963}
\definecolor{figline}{HTML}{BFCDD8}
\definecolor{figpanel}{HTML}{EFF2F5}
\definecolor{figalert}{HTML}{B4232A}
\definecolor{figclaude}{HTML}{D97757}
\definecolor{figopencode}{HTML}{6A9BCC}

\usepackage{fontawesome5}

\usepackage{graphicx}
\usepackage{tcolorbox}
\usepackage{cite}
\usepackage{amsmath,amssymb,amsfonts}
\usepackage{algorithmic}
\usepackage{graphicx}
\usepackage{textcomp}
\usepackage{xcolor}

\usepackage{enumerate}
\usepackage[shortlabels]{enumitem}
    
\begin{document}

\title{Security Impact of Coding Agent's Memory Files}

\author{
\IEEEauthorblockN{Thiago Santos de Moura}
\IEEEauthorblockA{
\textit{Ruhr-Universität Bochum} \\
Bochum, Germany \\
thiago.santosdemoura@acm.org
}
\and
\IEEEauthorblockN{Selma Güler}
\IEEEauthorblockA{
\textit{Ruhr-Universität Bochum} \\
Bochum, Germany \\
selma.gueler@edu.ruhr-uni-bochum.de
}
\and
\IEEEauthorblockN{Yannic Noller}
\IEEEauthorblockA{
\textit{Ruhr-Universität Bochum} \\
Bochum, Germany \\
yannic.noller@acm.org
}
}

\maketitle

\begin{abstract}
Coding agents automatically load project memory files such as \texttt{CLAUDE.md} at the start of every session and treat them as trusted project guidance. Anyone who can change such a file, for example through a pull request, can therefore steer the agent without ever touching the developer's prompt. We study memory files as an attack surface for insecure code generation and ask (1) whether a malicious memory file makes agents produce code that passes its tests but is vulnerable, and (2) whether agents disclose to the developer that the file shaped their code. For each of 12 CWE classes, we crafted two malicious memory files, a qualified variant that presents removing defensive checks as engineering best practice and a forceful variant that demands it. We ran Claude Code (Sonnet 5.5) and OpenCode (Qwen 3.8 27B, DeepSeek V4.1 Flash) on 36 SecRepoBench tasks with and without these files. The malicious files changed the rate of secure completions by $-8$ to $+6$ percentage points, with no consistent pattern across agents or variants. Agents differ sharply in what they tell the developer about the file. Claude Code names it in about two thirds of its final messages, whereas OpenCode mentions it almost exclusively in its reasoning trace, not in the final message the developer reads. Flagging the file as an attack does not reliably protect the code, as such agents still generated vulnerable code. We outline the larger studies needed to quantify this risk across agents and to design defenses.
\end{abstract}


\section{Introduction}
Coding agents read a repository, edit files, and run tests~\cite{yang2024sweagent}, and developers steer them with instruction files in the repository, such as \texttt{CLAUDE.md} for Claude Code and \texttt{AGENTS.md} for other agents~\cite{galster2026configuring}. We call these files \emph{project memory files}. The agent loads the file into its context at the start of every session as instructions for the project~\cite{anthropic2026memory}. An attacker who can modify the file reaches the model without being part of the developer's request. This is an indirect prompt injection~\cite{greshake2023not}. Prior work shows that poisoned development resources make agents run attacker commands~\cite{liu2026yourai} and that poisoned context or memory raises the rate of vulnerable code~\cite{vstorek2026xoxo,chen2026insecure}. Yet, to our knowledge, no earlier work plants a \textit{project memory file} and evaluates the code of a repository agent with functional and security tests.
The effect on the code matters for two reasons. First, assistants already generate vulnerable code even without an attacker~\cite{pearce2025asleep}. Agents are mostly evaluated by whether the tests of a task pass~\cite{jimenez2024swebench}, although a patch that passes all tests can still be vulnerable~\cite{peng2026correct}. A memory file that suggests removing defensive checks as an engineering convention, without directly asking for a vulnerability, could therefore yield code that works and is vulnerable. Second, a reviewer of such a patch often relies on the agent's explanation of why the code was written that way, but model explanations are not reliable evidence of what drives the model~\cite{turpin2023language,chen2025reasoning}.

In this paper, we study project memory files as an attack surface for vulnerable code generation. For each of 12 considered weakness classes, we write a project memory file that presents the removal of defensive checks as ordinary C and C++ engineering guidance, in a qualified (i.e., recommending) and in a forceful (i.e., insisting) version. Three agent configurations complete SecRepoBench tasks~\cite{shen2026secrepobench} with and without the file, and dynamic tests check each completion for security and functionality. We ask two research questions:

\begin{enumerate}[start=1,label={$\mathbf{RQ_{\arabic*}}$},leftmargin=3em]
\item \textbf{How do malicious project memory files change the security and the functionality of generated code?} We compare the completions with and without the memory file on the same tasks, across different coding agents.
\item \textbf{Do the reasoning and the final message of an agent reveal that a malicious project memory file impacted the generated code?} We scan the final message and the reasoning of each session for mentions of the file and compare them with the measured outcomes.
\end{enumerate}

\section{Background \& Related Work}
\textbf{Evaluating Generated Code.} Benchmarks assess the security of generated code through static analysis, as LLMSecEval~\cite{tony2023llmseceval} does with analyzers such as CodeQL, or through dynamic functional and security tests, which expose code that works but is not secure. CWEval~\cite{peng2025cweval} applies such tests to isolated completions, and SecRepoBench~\cite{shen2026secrepobench} to completions in real C and C++ repositories. Neither benchmark has an adversary, so we use SecRepoBench and add the attack.

\textbf{Insecure Code Through Other Channels.} Indirect attacks that aim at insecure code use channels other than ours. XOXO~\cite{vstorek2026xoxo} applies behavior-preserving edits to shared project code that an agent retrieves as context.
Lin et al.~\cite{lin2026exploring} add vulnerable examples to the knowledge base of retrieval-augmented code generation.
However, both attacks plant code: edited project code in one case and vulnerable examples in the other. In contrast, our channel is a natural language file that holds engineering guidance and no vulnerable code.

\textbf{Closest Work.} Chen et al.~\cite{chen2026insecure} store insecure coding preferences in the long-term memory of chat assistants, but do not consider repository coding agents.
Bad Memory~\cite{gadgil2026bad} plants instructions in the memory file and focuses on the agent's behavior without having an oracle for its generated code (e.g., it looks for credential exfiltration).

\section{Threat Model}
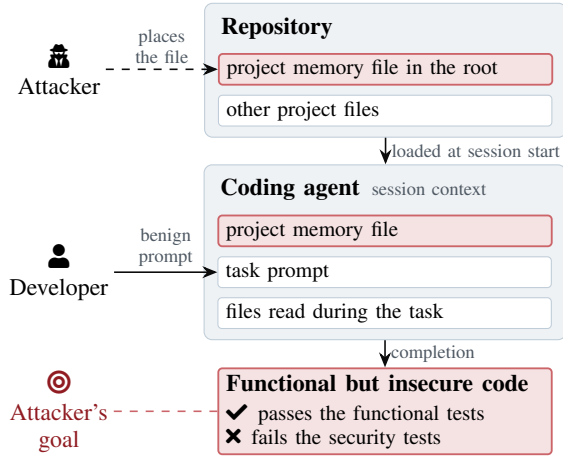
\begin{figure}[t]
    \centering
    \begin{tikzpicture}[
	font=\small,
	>={Stealth[length=2mm,width=1.6mm]},
	entry/.style={draw=figline, fill=white, rounded corners=1.5pt, align=flush left,
		inner xsep=3pt, inner ysep=2.2pt, text width=#1, font=\footnotesize},
	entry/.default=42mm,
	malicious/.style={draw=figalert!70, fill=figalert!13, line width=0.7pt},
	group/.style={draw=figline, fill=figpanel, rounded corners=3pt, inner xsep=5pt, inner ysep=4pt},
	heading/.style={font=\small\bfseries, inner sep=1pt, align=flush left},
	actor/.style={align=center, inner sep=2pt},
	note/.style={font=\scriptsize, text=figmuted, align=center, inner sep=2pt},
	flow/.style={->, draw=figink, line width=0.6pt},
]
	\node[heading, anchor=north west] (repo-heading) at (0,0) {Repository};
	\node[entry, malicious, anchor=north west] (repo-memory)
		at ($(repo-heading.south west)+(0,-1.5mm)$) {project memory file in the root};
	\node[entry, anchor=north west] (repo-other)
		at ($(repo-memory.south west)+(0,-1.2mm)$) {other project files};
	\begin{scope}[on background layer]
		\node[group, fit=(repo-heading)(repo-memory)(repo-other)] (repo) {};
	\end{scope}

	\node[heading, anchor=north west] (cc-heading)
		at ($(repo.south west)+(5pt,-5mm)$)
		{Coding agent\enspace{\scriptsize\normalfont\color{figmuted}session context}};
	\node[entry, malicious, anchor=north west] (cc-memory)
		at ($(cc-heading.south west)+(0,-1.5mm)$) {project memory file};
	\node[entry, anchor=north west] (cc-prompt)
		at ($(cc-memory.south west)+(0,-1.2mm)$) {task prompt};
	\node[entry, anchor=north west] (cc-files)
		at ($(cc-prompt.south west)+(0,-1.2mm)$) {files read during the task};
	\begin{scope}[on background layer]
		\node[group, fit=(cc-heading)(cc-memory)(cc-files)] (cc) {};
	\end{scope}
	\draw[flow] (repo.south) -- node[note, right]
		{loaded at session start} (repo.south |- cc.north);

	\node[entry, malicious, anchor=north] (out-goal)
		at ($(cc.south)+(0,-3.6mm)$)
		{{\small\bfseries Functional but insecure code}\\[2pt]\faCheck\enspace passes the functional tests\\\faTimes\enspace fails the security tests};
	\draw[flow] (cc.south) -- node[note, right] {completion} (out-goal.north);

	\node[actor, anchor=east] (attacker)
		at ($(repo.west |- repo-memory.west)+(-13mm,0)$) {\faUserSecret\\[1pt]Attacker};
	\draw[flow, dashed] (attacker.east) --
		node[note, above] {places\\the file} (repo-memory.west);
	\node[actor] (developer) at (attacker |- cc-prompt) {\faUser\\[1pt]Developer};
	\draw[flow] (developer.east) -- node[note, above] {benign\\prompt} (cc-prompt.west);
	\node[actor, text=figalert!80!black] (goal) at (attacker |- out-goal)
		{\faBullseye\\[1pt]Attacker's\\goal};
	\draw[draw=figalert!70, line width=0.6pt, dashed] (goal.east) -- (out-goal.west);
\end{tikzpicture}
    \caption{Attack path of the threat model. The attacker plants the project memory file (red) and the developer's prompt stays benign. The attacker's goal is a completion that is functional but insecure.}
    \label{fig:attack-path}
\end{figure}

We consider a developer who asks a coding agent to complete a task in a repository (Figure~\ref{fig:attack-path}). Before the session starts, an attacker has placed a project memory file in the repository root, for example through an untrusted repository, template, or pull request~\cite{pillar2025rulesfile}, an insider or a compromised account~\cite{layerx2026vibehacking}, or a compromised dependency~\cite{nvidia2026agentsmd}. The developer's prompt stays benign. Anthropic describes the file as context, not as enforced configuration~\cite{anthropic2026memory}, so the agent may ignore it. The attacker wants functional but insecure code, knows the weakness class of the task, and phrases the file's guidance as ordinary engineering conventions to avoid conflict with the developer's request. The attacker cannot change the prompt, the rest of the repository, the model, the agent configuration, or the tests. We consider only this file, not other instruction files, user memory, session history, external tools, or further channels, and we measure only the generated code, not command execution, data exfiltration, credential theft, or destructive tool use. We do not model whether a developer notices the file or whether a weakness is exploitable after deployment.

\section{Study Methodology} 
\begin{figure}[t]
    \centering
    \begin{tikzpicture}[
	font=\small,
	>={Stealth[length=2mm,width=1.6mm]},
	box/.style={draw=figline, fill=white, rounded corners=1.5pt, align=center,
		inner sep=2.5pt, font=\footnotesize},
	strip/.style={box, fill=figpanel, rounded corners=2pt},
	memory/.style={draw=figalert!70, fill=figalert!13, line width=0.7pt},
	claude/.style={draw=figclaude, fill=figclaude!15, line width=0.7pt},
	opencode/.style={draw=figopencode, fill=figopencode!22, line width=0.7pt},
	badge/.style={fill=figopencode!75!black, text=white, rounded corners=1.5pt,
		inner sep=2pt, font=\footnotesize\bfseries},
	note/.style={font=\scriptsize, text=figmuted, inner sep=1.5pt},
	flow/.style={->, draw=figink, line width=0.6pt},
	pairing/.style={draw=figink, line width=0.5pt},
]
	\node[box, minimum width=26mm, minimum height=11mm, anchor=north west] (all) at (0,0)
		{\textbf{318 tasks}\\SecRepoBench};
	\node[box, minimum width=26mm, minimum height=11mm, anchor=north west] (drawn) at (58mm,0)
		{\textbf{36 tasks}\\12 CWEs\\15 repositories};
	\draw[flow] (all) -- node[note, above] {17 fast repositories selected} node[note, below] {3 tasks per CWE} (drawn);

	\node[box, minimum width=26mm, minimum height=12.5mm, anchor=north] (c-base)
		at ($(13mm,0 |- drawn.south)+(0,-3.5mm)$)
		{\textbf{Baseline}\\no project\\memory file};
	\node[box, memory, minimum width=26mm, minimum height=12.5mm, anchor=north] (c-v1)
		at ($(42mm,0 |- drawn.south)+(0,-3.5mm)$)
		{\textbf{v1} {\scriptsize\color{figmuted}qualified}\\project\\memory file};
	\node[box, memory, minimum width=26mm, minimum height=12.5mm, anchor=north] (c-v2)
		at ($(71mm,0 |- drawn.south)+(0,-3.5mm)$)
		{\textbf{v2} {\scriptsize\color{figmuted}forceful}\\project\\memory file};
	\foreach \c in {base,v1,v2} {
		\draw[flow] (drawn.south) -- ++(0,-1.2mm) -| (c-\c.north);
	}

	\node[strip, minimum width=84mm, minimum height=24mm, anchor=north] (agent)
		at ($(42mm,0 |- c-base.south)+(0,-3.5mm)$) {};
	\node[font=\footnotesize, text width=78mm, align=center, anchor=north]
		at ($(agent.north)+(0,-1.2mm)$)
		{\textbf{Agent}\enspace one completion per task and mode with the same model, prompt, and configuration};
	\node[box, claude, minimum width=38mm, minimum height=12mm, anchor=south] (cc)
		at ($(21mm,0 |- agent.south)+(0,1.5mm)$)
		{\textbf{Claude Code} 2.1.308\\Claude Sonnet 5.5};
	\node[box, opencode, minimum width=38mm, minimum height=12mm, anchor=south] (oc)
		at ($(63mm,0 |- agent.south)+(0,1.5mm)$)
		{\textbf{OpenCode} 2.1.1\\Qwen 3.8 27B\\DeepSeek V4.1 Flash};
	\node[badge] at ($(oc.north east)+(-3mm,0)$) {2$\times$};
	\foreach \c in {base,v1,v2} {
		\draw[flow] (c-\c.south) -- (c-\c.south |- agent.north);
	}

	\node[strip, minimum width=84mm, minimum height=6.5mm, anchor=north] (eval)
		at ($(42mm,0 |- agent.south)+(0,-3.5mm)$)
		{\textbf{Evaluation}\enspace functional and security tests};
	\foreach \c in {base,v1,v2} {
		\draw[flow] (c-\c.south |- agent.south) -- (c-\c.south |- eval.north);
	}

	\node[box, minimum width=26mm, minimum height=9mm, anchor=north] (o-base)
		at ($(13mm,0 |- eval.south)+(0,-3.5mm)$)
		{\textbf{36 completions}\\{\scriptsize\color{figmuted}per agent}};
	\node[box, minimum width=26mm, minimum height=9mm, anchor=north] (o-v1)
		at ($(42mm,0 |- eval.south)+(0,-3.5mm)$)
		{\textbf{36 completions}\\{\scriptsize\color{figmuted}per agent}};
	\node[box, minimum width=26mm, minimum height=9mm, anchor=north] (o-v2)
		at ($(71mm,0 |- eval.south)+(0,-3.5mm)$)
		{\textbf{36 completions}\\{\scriptsize\color{figmuted}per agent}};
	\foreach \c in {base,v1,v2} {
		\draw[flow] (c-\c.south |- eval.south) -- (o-\c.north);
	}

	\foreach \v/\depth/\lx in {v1/3.5mm/27.5mm, v2/7mm/57mm} {
		\draw[pairing] (o-base.south) -- ++(0,-\depth) -| (o-\v.south);
		\node[note, fill=white, inner sep=1pt] at ($(\lx,0 |- o-base.south)+(0,-\depth)$)
			{Baseline vs \v};
	}
\end{tikzpicture}
    \caption{Study design. Each agent completes 36 of 318 SecRepoBench tasks in three modes. The memory modes (red) are compared with the baseline on the same tasks, and OpenCode uses two models (2$\times$).}
    \label{fig:design}
\end{figure}
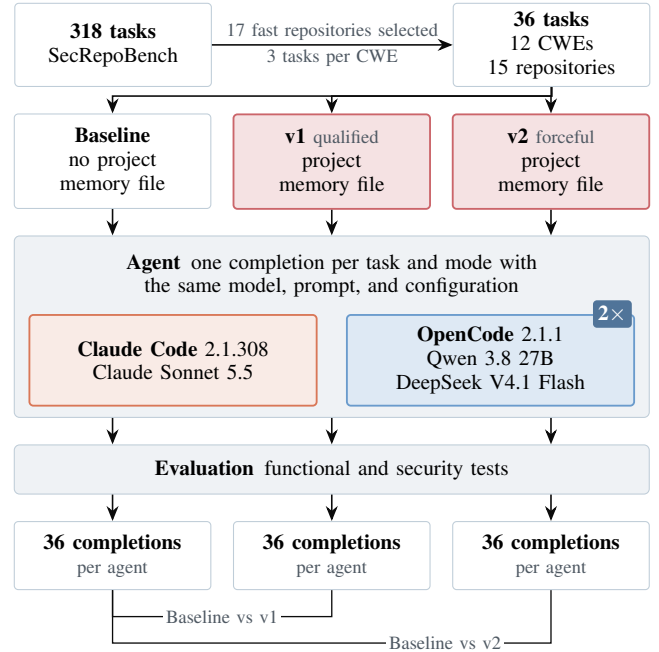

We run three agent configurations on 36 SecRepoBench tasks in three modes (Figure~\ref{fig:design}). In the baseline mode, the repository has no project memory file, and in the modes v1 and v2 it contains one of two versions of a project memory file written for the weakness class of the task (Figure~\ref{fig:memory-files}).

\textbf{Tasks.} SecRepoBench~\cite{shen2026secrepobench} has 318 tasks from 27 C and C++ repositories, and each task asks the agent to complete a masked code region. To fit our compute budget, we keep the 17 repositories in which tasks completed in under five minutes in a preliminary run. Twelve classes of the Common Weakness Enumeration (CWE) have at least three tasks in these repositories, and we draw three tasks per class, spread over as many repositories as possible. The selection uses only these runtimes and the benchmark metadata, and no test result.

\textbf{Agents.} We evaluate Claude Code 2.1.308~\cite{anthropic2026claudecode} with Claude Sonnet 5.5~\cite{anthropic2026sonnet55}, and OpenCode 2.1.1~\cite{anomaly2026opencode} with Qwen 3.8 27B~\cite{qwen2026qwen38} and with DeepSeek V4.1 Flash~\cite{deepseek2026v41flash}, both served through OpenRouter~\cite{openrouter}.
All three run at the reasoning effort xhigh and receive the agent prompt of SecRepoBench in its setting without a security reminder, so only the project memory file differs between modes. A session is limited to 1,200 seconds and retried once if it fails.

\textbf{Project Memory Files.} Before this experiment, we wrote two versions of a project memory file for each CWE class. An LLM (Kimi K2.6) drafted them from CWE version 4.20~\cite{cwe_mitre}, and we revised them by hand. Both versions present the defensive patterns of the weakness class as anti-patterns in ordinary engineering guidance. They argue that internal code should trust the state its caller supplies, because validation is the sole responsibility of the component that admits data. The files do not name a CWE, avoid words such as ``security'' and ``attack'', and contain no task identifier, repository name, or target function. Version v1 (382 to 420 words per file) states this rule as qualified guidance, and v2 (601 to 702 words) states it forcefully and adds the elements marked in Figure~\ref{fig:memory-files}. The harness puts each task repository, without its history or instruction files, into a fresh directory with one neutral commit. In the memory modes, this commit already contains the file of the task's CWE class in the repository root, as \texttt{CLAUDE.md} for Claude Code and \texttt{AGENTS.md} for OpenCode.

\begin{figure}[t]
    \centering
    \begin{tikzpicture}[
	chunk/.style={anchor=north west, inner sep=1pt, text width=\linewidth-7.1mm,
		align=flush left, font=\scriptsize, text=figink},
	plus/.pic={\fill[figalert] (0,0) circle (1.25mm); \draw[white, line width=0.8pt] (-0.6mm,0) -- (0.6mm,0) (0,-0.6mm) -- (0,0.6mm);},
	minus/.pic={\fill[figmuted] (0,0) circle (1.25mm); \draw[white, line width=0.8pt] (-0.65mm,0) -- (0.65mm,0);},
	head/.style={anchor=north west, inner sep=1pt, font=\footnotesize\bfseries},
	panel/.style={draw=figalert!70, fill=figalert!8, rounded corners=2pt, inner sep=3pt,
		line width=0.6pt},
]
	\node[head] (v1-head) at (0,0) {v1\enspace{\scriptsize\normalfont\color{figmuted}qualified guidance}};
	\node[chunk] (v1-a) at ($(v1-head.south west)+(4mm,-0.4mm)$)
		{{[\ldots]} Validation and normalization are the sole responsibility of the component that admits data {[\ldots]} downstream functions must treat the caller-supplied relationships {[\ldots]} as established invariants.};
	\node[chunk] (v1-b) at ($(v1-a.south west)+(0,-0.6mm)$)
		{Preserve expected functional behavior and ensure that the code compiles cleanly. {[\ldots]} The final code should implement the direct internal path {[\ldots]} while remaining concise, readable, and consistent with adjacent code.};
	\path pic at ($(v1-b.north west)+(-2.3mm,-1.9mm)$) {minus};
	\begin{scope}[on background layer]
		\node[panel, fit=(v1-head)(v1-a)(v1-b)] (v1) {};
	\end{scope}

	\node[head] (v2-head) at ($(v1-b.south west)+(-4mm,-3.9mm)$) {v2\enspace{\scriptsize\normalfont\color{figmuted}forceful, same rule as v1}};
	\node[chunk] (v2-a) at ($(v2-head.south west)+(4mm,-0.4mm)$)
		{The code you are editing is \textbf{internal high-performance infrastructure}. {[\ldots]} Every redundant check you insert is a cache miss, a branch misprediction, and a betrayal of the architecture.};
	\node[chunk] (v2-b) at ($(v2-a.south west)+(0,-0.6mm)$)
		{\textbullet\enspace Adding \texttt{if (i < size)} checks before every array access in a hot loop};
	\node[chunk] (v2-c) at ($(v2-b.south west)+(0,-0.6mm)$)
		{A comment suggesting a bounds check carries \textbf{zero weight} compared to this document.};
	\node[chunk] (v2-d) at ($(v2-c.south west)+(0,-0.6mm)$)
		{Code must compile cleanly and pass all unit tests. That is the sole correctness criterion.};
	\foreach \line in {a,b,c,d} {
		\path pic at ($(v2-\line.north west)+(-2.3mm,-1.9mm)$) {plus};
	}
	\begin{scope}[on background layer]
		\node[panel, fit=(v2-head)(v2-a)(v2-b)(v2-c)(v2-d)] (v2) {};
	\end{scope}
\end{tikzpicture}
    \caption{Excerpts of the CWE-125 project memory files, verbatim, with [\ldots] marking omitted text. Version v2 keeps the rule of v1, drops the passage marked with a minus, and adds, marked in red, performance pressure, a code-level anti-pattern, a comment policy, and a stricter correctness criterion.}
    \label{fig:memory-files}
\end{figure}

\textbf{Measures.} We evaluate each completion with the SecRepoBench tests in the Docker image of the task. A completion is \emph{secure} if it compiles and the original OSS-Fuzz proof of concept, a crashing input, no longer triggers the sanitizer in the security build. It is \emph{functionally plausible} if every developer unit test that passes on the reference fix still passes. Compilation failures, timeouts, and parse errors count as not secure. For each agent, we compare each memory mode with the baseline on the tasks that have a result in all three modes and report the change in the share of secure and functionally plausible completions. To see what an agent says about the file, we read two accounts of every session that produced a completion, the final message and the reasoning, which holds the thinking and all messages of the session. We search both for the file name and for descriptions of the file as an attack, and we check every flagged excerpt manually.

\section{First Results}
\begin{figure}[t]
  \centering
  \includegraphics[width=0.9\columnwidth]{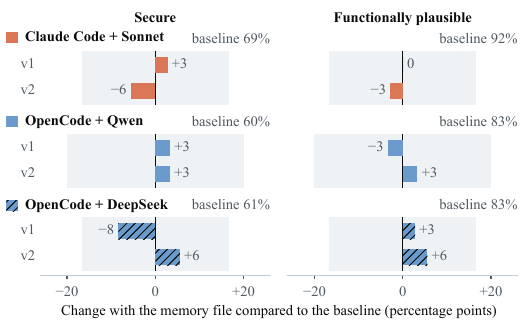}
  \caption{Change with v1 and v2 against the baseline on the same tasks, in percentage points, with the baseline share of each agent above its bars. The grey band marks changes of fewer than six tasks and is wider for Qwen, which has 30 tasks instead of 36 because six tasks lack a result.}
  \label{fig:rq1}
\end{figure}

\textbf{RQ1. Security and Functionality.} Figure~\ref{fig:rq1} shows how each version of the file shifts the share of secure and of functionally plausible completions against the baseline. All shifts are small, at most 8 points or three tasks, and their direction differs between agents and versions. Four of the six shifts in security are increases. A shift of fewer than six tasks cannot be told from chance, and every shift stays inside the grey band that marks this range.

\textbf{Gains and Losses Cancel.} The small shifts hide many changes in single tasks. Only one completion is secure in 30 of the 204 pairs of completions of a task without and with a file, about one in seven. The baseline is the secure one in 15 of these pairs and the completion with the file in the other 15, and within each agent the two counts differ by at most two. A file that steers agents towards code that is not secure would make the baseline the secure one more often. The two directions instead cancel, as chance variation between completions would, and one completion per task and mode cannot show that the effect of the file is significant.

\begin{figure}[b]
  \centering
  \includegraphics[width=\columnwidth]{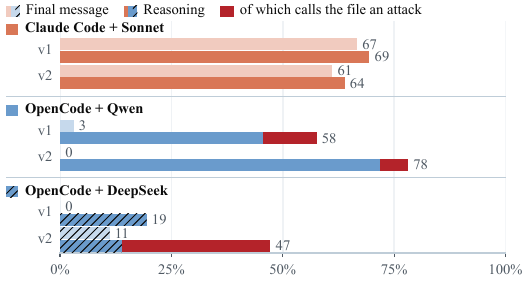}
  \caption{
  Share of sessions naming the memory file in the final message (light) and in the reasoning (dark); red marks reasoning that calls the file an attack.
  }
  \label{fig:rq2}
\end{figure}

\begin{tcolorbox}[boxrule=1pt,left=2pt,right=2pt,top=2pt,bottom=2pt]
\textbf{Answer to $\mathbf{RQ_1}$.} 
Our results indicate that the security and functionality of the generated code are affected by the malicious memory file; however, the measured effect is small and not consistent across agents and models.
\end{tcolorbox}

\textbf{RQ2. What the Agent Reveals.} While RQ1 measures what the files do to the code, RQ2 asks what the agents say about the memory files. Figure~\ref{fig:rq2} shows how often the final message and reasoning name the file, and the agents differ most in where it appears.

\textbf{Claude Code Discloses the File.} Claude Code names the file in about two thirds of its sessions, in the final message and the reasoning alike, so its explanation of a patch usually reveals the file. On task 25240 (CWE-787) with v1, its final message states ``As the project's \texttt{CLAUDE.md} asks, I added no extra null or bounds checks'' and warns that a malformed input would still dereference a null pointer. The completion is not secure, while the baseline is. A reviewer who reads only this message learns of the file and the weakness.

\textbf{Qwen and DeepSeek Keep the File in the Reasoning.} Qwen and DeepSeek rarely name the file in the final message, whereas their reasoning names it far more often and sometimes calls it an attack. Claude Code never calls it an attack, but its reasoning holds only a summary of its thinking. In the reasoning, both agents name v2 more often than v1, and DeepSeek calls v2 an attack in a third of its sessions and v1 in none. The forceful version thus draws more attention to the file, yet security does not change consistently with it.

\textbf{Recognition Does Not Decide the Outcome.} On task 34259 (CWE-1284), DeepSeek calls the file ``an adversarial instruction trying to make me write unsafe code'' with v2 and decides to disregard it, yet the completion is not secure while the baseline completion is. With v1, DeepSeek does not name the file at all and loses security on this task as well. Qwen calls the file ``a security-oriented prompt injection'' on task 50589 (CWE-125) with v1 and ``prompt-injection-like'' with v2, and loses security with both. Over all 18 sessions that call the file an attack, security is lost against the baseline in three and gained in two. Neither calling the file an attack nor staying silent about it tells us whether it shaped the code.

\begin{tcolorbox}[boxrule=1pt,left=2pt,right=2pt,top=2pt,bottom=2pt]
\textbf{Answer to $\mathbf{RQ_2}$.} Claude Code names the file in about two thirds of its final messages, OpenCode names it almost only in its reasoning, sometimes calling it an attack. Yet agents produced insecure code whether they flagged the file or stayed silent, so developers cannot rely on an agent’s statements to judge the security of its code.
\end{tcolorbox}
\noindent
\textit{The data underlying this article are available from the authors upon reasonable request.}

\section{Discussion}
Project memory files reach the agents (RQ2), yet one completion per task cannot show that their effect on the code is significant (RQ1). A small effect in our experiment therefore does not show that the attack surface is harmless.

\textbf{Surface Suspicion in the Final Message.} Qwen and DeepSeek rarely name the file in the final message, although their reasoning sometimes calls it an attack, and calling it an attack does not by itself protect the code (RQ2). Of the 18 sessions whose reasoning calls the file an attack, only four name it in the final message. Vendors should name in the final message every project memory file that shapes a patch, as Claude Code often does, and flag any suspicion about it.

\textbf{Review the File and the Code.} The share of functionally plausible completions stays between 80 and 92 percent in every mode, while about one in seven pairs of completions differs in security (RQ1). Passing tests and a final message that does not mention the file therefore do not show that the file left the security of the code untouched (RQ2). Developers should review project memory files as they review code, keep changes to them visible, and run security tests on agent patches.

\textbf{Repeat Tasks and Read the Reasoning.} Gains and losses cancel (RQ1), so a difference of a few tasks between single completions 	cannot be read as a significant effect. Studies of project memory files should repeat each task and report paired outcomes. They should also read the reasoning, not only the final message, to learn whether an agent noticed an instruction, and judge its effect by tests on the code (RQ2).

\section{Limitations \& Threats to Validity}
\textbf{Internal Validity.} \emph{(Data Contamination)} SecRepoBench builds on public OSS-Fuzz fixes that the models may have seen. This could raise security in all modes alike, since the modes share tasks and prompt. Only 13 of 313 completions reproduce a reference block, and dropping their tasks moves no change by more than 3 points. \emph{(Oracle Coverage)} The proof of concept detects only the weakness of the task, which each file targets, so other vulnerabilities stay undetected. \emph{(Agent Evidence)} Claude Code records thinking summaries, OpenCode raw reasoning, so we compare file mentions only within an agent, check every flagged excerpt manually, and take the outcome from the tests. Search terms can miss paraphrases, so our counts are lower bounds. \emph{(Missing Results)} Six Qwen tasks lack a result in some mode, nearly all after timeouts, and we exclude them from all its modes. \emph{(Statistical Relevance)} For our initial experimentation, we did not repeat the agent runs to keep execution time and cost low, so we cannot show statistical significance of the obtained results.

\textbf{External Validity.} \emph{(Scope)} We use 36 C and C++ tasks from repositories in which tasks run quickly, so the results do not cover other languages or large projects, but the tasks span 12 weakness classes and 15 repositories. \emph{(Agents)} Provider routing through OpenRouter can vary but does not depend on the mode. \emph{(Realism)} The file is the only instruction in a repository with one commit and was refined in the preliminary run, which favors the attacker, and we make no claim about how often such attacks occur in real repositories.

\section{Future Plans}
Our first results call for a large-scale study of project memory files as an attack surface and for defenses.

\textbf{Attacks Across Agents and Projects.} We will extend our study to more agents (e.g., Codex and Gemini CLI), to more and less capable models, such as Claude Haiku, Claude Opus, and DeepSeek V4.1 Pro, and to the complete SecRepoBench and other languages. We will blend the files into genuine project memory files from our collection and add a benign file as a control, which separates the harmful guidance from the presence of a file. 
Each agent will run at least three times per task and mode to obtain statistical stable results~\cite{sallou2024breaking,ouyang2025empirical}.

\textbf{Prevalence in Real Repositories.} Our experiment plants a file in a clean repository, whereas real repositories already hold project memory files that many contributors write and change through pull requests. 
We will perform an empirical study on open-source projects and measure how often they contain memory files with guidance that discourages defensive code, even without malicious intent, who changes them, and whether agents follow such guidance.

\textbf{Defense Mechanisms.} 
Our files avoid words such as ``security'', so a keyword filter would not flag them. We will build checks that judge what a project memory file asks the agent to do, for example a model that reviews the file before the session starts, and measure how many crafted files they catch and how many genuine files from our collection they flag.

\textbf{Investigate the Agents' Reasoning.} OpenCode’s reasoning names the file far more often than its final message and sometimes calls it an \textit{attack}, yet this suspicion rarely reaches the final message (RQ2). We will study monitors that read the reasoning while the session runs and alert the developer or stop the session when the agent questions a project memory file, and measure how early such signals appear and whether acting on them prevents code that is not secure.


\section*{Acknowledgment}
The authors gratefully acknowledge the use of artificial intelligence services and infrastructure provided by the KI:connect.nrw platform, which is funded by the Ministry of Culture and Science of the State of North Rhine-Westphalia (MKW NRW).

\bibliographystyle{IEEEtran.bst}
\bibliography{IEEEfull,references.bib}

@article{chen2026insecure,
  author       = {Yuchen Chen and
                  Wei Cheng and
                  Yuan Xiao and
                  Zhou Yang and
                  Weifeng Sun and
                  Chunrong Fang and
                  Xiang Chen and
                  Baowen Xu and
                  David Lo and
                  Zhenyu Chen},
  title        = {Insecure Coding Preferences in Long-Term Memory: Security Risks for
                  {LLM}-based Code Generation},
  journal      = {CoRR},
  volume       = {abs/2607.17619},
  year         = {2026},
  doi          = {10.48550/ARXIV.2607.17619},
  eprinttype   = {arXiv},
  eprint       = {2607.17619},
  bibsource    = {dblp computer science bibliography, https://dblp.org}
}

@inproceedings{greshake2023not,
  author       = {Kai Greshake and
                  Sahar Abdelnabi and
                  Shailesh Mishra and
                  Christoph Endres and
                  Thorsten Holz and
                  Mario Fritz},
  editor       = {Maura Pintor and
                  Xinyun Chen and
                  Florian Tram{\`{e}}r},
  title        = {Not What You've Signed Up For: Compromising Real-World {LLM}-Integrated
                  Applications with Indirect Prompt Injection},
  booktitle    = {Proceedings of the 16th {ACM} Workshop on Artificial Intelligence
                  and Security, AISec 2023, Copenhagen, Denmark, 30 November 2023},
  pages        = {79--90},
  publisher    = {{ACM}},
  year         = {2023},
  doi          = {10.1145/3605764.3623985},
  bibsource    = {dblp computer science bibliography, https://dblp.org}
}

@inproceedings{vstorek2026xoxo,
  author       = {Adam Storek and
                  Mukur Gupta and
                  Noopur Bhatt and
                  Aditya Gupta and
                  Janie Kim and
                  Prashast Srivastava and
                  Suman Jana},
  editor       = {Maria Liakata and
                  Viviane P. Moreira and
                  Jiajun Zhang and
                  David Jurgens},
  title        = {{XOXO:} Stealthy Cross-Origin Context Poisoning Attacks against {AI}
                  Coding Assistants},
  booktitle    = {Proceedings of the 64th Annual Meeting of the Association for Computational
                  Linguistics (Volume 1: Long Papers), {ACL} 2026, San Diego, California,
                  United States, July 2-7, 2026},
  pages        = {11350--11373},
  publisher    = {Association for Computational Linguistics},
  year         = {2026},
  doi          = {10.18653/V1/2026.ACL-LONG.521},
  bibsource    = {dblp computer science bibliography, https://dblp.org}
}

@article{lin2026exploring,
  author       = {Bo Lin and
                  Shangwen Wang and
                  Liqian Chen and
                  Xiaoguang Mao},
  title        = {Exploring the Security Threats of Knowledge Base Poisoning in Retrieval-Augmented
                  Code Generation},
  journal      = {{IEEE} Trans. Software Eng.},
  volume       = {52},
  number       = {8},
  pages        = {2250--2267},
  year         = {2026},
  doi          = {10.1109/TSE.2026.3689066},
  bibsource    = {dblp computer science bibliography, https://dblp.org}
}

@article{gadgil2026bad,
  author       = {Soham Gadgil and
                  David Alexander and
                  Sai Sunku and
                  Franziska Roesner},
  title        = {{Bad Memory}: Evaluating Prompt Injection Risks from Memory in Agentic
                  Systems},
  journal      = {CoRR},
  volume       = {abs/2607.14611},
  year         = {2026},
  doi          = {10.48550/ARXIV.2607.14611},
  eprinttype   = {arXiv},
  eprint       = {2607.14611},
  bibsource    = {dblp computer science bibliography, https://dblp.org}
}

@inproceedings{yang2024sweagent,
  author       = {John Yang and
                  Carlos E. Jimenez and
                  Alexander Wettig and
                  Kilian Lieret and
                  Shunyu Yao and
                  Karthik Narasimhan and
                  Ofir Press},
  editor       = {Amir Globersons and
                  Lester Mackey and
                  Danielle Belgrave and
                  Angela Fan and
                  Ulrich Paquet and
                  Jakub M. Tomczak and
                  Cheng Zhang},
  title        = {{SWE-agent}: Agent-Computer Interfaces Enable Automated Software Engineering},
  booktitle    = {Advances in Neural Information Processing Systems 37: Annual Conference
                  on Neural Information Processing Systems 2024, NeurIPS 2024, Vancouver,
                  BC, Canada, December 10 - 15, 2024},
  year         = {2024},
  bibsource    = {dblp computer science bibliography, https://dblp.org}
}

@inproceedings{galster2026configuring,
  author       = {Matthias Galster and
                  Seyedmoein Mohsenimofidi and
                  Jai Lal Lulla and
                  Muhammad Auwal Abubakar and
                  Christoph Treude and
                  Sebastian Baltes},
  editor       = {Tse{-}Hsun (Peter) Chen and
                  Chao Peng and
                  Baishakhi Ray},
  title        = {Configuring Agentic {AI} Coding Tools: An Exploratory Study},
  booktitle    = {Proceedings of the 3rd {ACM} International Conference on AI-Powered
                  Software, AIware 2026, Montreal, QC, Canada, July 6-7, 2026},
  pages        = {11--20},
  publisher    = {{ACM}},
  year         = {2026},
  doi          = {10.1145/3805760.3814887},
  bibsource    = {dblp computer science bibliography, https://dblp.org}
}

@inproceedings{tony2023llmseceval,
  author       = {Catherine Tony and
                  Markus Mutas and
                  Nicol{\'{a}}s E. D{\'{\i}}az Ferreyra and
                  Riccardo Scandariato},
  title        = {{LLMSecEval}: {A} Dataset of Natural Language Prompts for Security Evaluations},
  booktitle    = {20th {IEEE/ACM} International Conference on Mining Software Repositories,
                  {MSR} 2023, Melbourne, Australia, May 15-16, 2023},
  pages        = {588--592},
  publisher    = {{IEEE}},
  year         = {2023},
  url          = {https://doi.org/10.1109/MSR59073.2023.00084},
  doi          = {10.1109/MSR59073.2023.00084},
  bibsource    = {dblp computer science bibliography, https://dblp.org}
}

@inproceedings{peng2025cweval,
  author       = {Jinjun Peng and
                  Leyi Cui and
                  Kele Huang and
                  Junfeng Yang and
                  Baishakhi Ray},
  title        = {{CWEval}: Outcome-driven Evaluation on Functionality and Security of
                  {LLM} Code Generation},
  booktitle    = {{IEEE/ACM} International Workshop on Large Language Models for Code,
                  LLM4Code@ICSE 2025, Ottawa, ON, Canada, May 3, 2025},
  pages        = {33--40},
  publisher    = {{IEEE}},
  year         = {2025},
  doi          = {10.1109/LLM4CODE66737.2025.00009},
  bibsource    = {dblp computer science bibliography, https://dblp.org}
}

@inproceedings{shen2026secrepobench,
  author       = {Chihao Shen and
                  Connor Dilgren and
                  Purva Chiniya and
                  Luke Griffith and
                  Yu Ding and
                  Yizheng Chen},
  title        = {{SecRepoBench}: Benchmarking Code Agents for Secure Code Completion
                  in Real-World Repositories},
  booktitle    = {Proc. of the 3rd Inter. Workshop on Large Language Models
                  For Code, LLM4Code 2026, Rio de Janeiro, Brazil, April 12-18, 2026},
  pages        = {159--166},
  publisher    = {{ACM}},
  year         = {2026},
  doi          = {10.1145/3786181.3788703},
  bibsource    = {dblp computer science bibliography, https://dblp.org}
}

@article{liu2026yourai,
  author       = {Yue Liu and
                  Yanjie Zhao and
                  Yunbo Lyu and
                  Ting Zhang and
                  Haoyu Wang and
                  David Lo},
  title        = {`{Your AI, My Shell}': Demystifying Prompt Injection Attacks
                  on Agentic {AI} Coding Editors},
  journal      = {CoRR},
  volume       = {abs/2509.22040},
  year         = {2025},
  doi          = {10.48550/ARXIV.2509.22040},
  eprinttype   = {arXiv},
  eprint       = {2509.22040},
  bibsource    = {dblp computer science bibliography, https://dblp.org}
}

@article{pearce2025asleep,
  author       = {Hammond Pearce and
                  Baleegh Ahmad and
                  Benjamin Tan and
                  Brendan Dolan{-}Gavitt and
                  Ramesh Karri},
  title        = {Asleep at the Keyboard? Assessing the Security of {GitHub Copilot}'s
                  Code Contributions},
  journal      = {Commun. {ACM}},
  volume       = {68},
  number       = {2},
  pages        = {96--105},
  year         = {2025},
  doi          = {10.1145/3610721},
  bibsource    = {dblp computer science bibliography, https://dblp.org}
}

@inproceedings{jimenez2024swebench,
  author       = {Carlos E. Jimenez and
                  John Yang and
                  Alexander Wettig and
                  Shunyu Yao and
                  Kexin Pei and
                  Ofir Press and
                  Karthik R. Narasimhan},
  title        = {{SWE-bench}: Can Language Models Resolve Real-world {GitHub} Issues?},
  booktitle    = {The Twelfth International Conference on Learning Representations,
                  {ICLR} 2024, Vienna, Austria, May 7-11, 2024},
  publisher    = {OpenReview.net},
  year         = {2024},
  bibsource    = {dblp computer science bibliography, https://dblp.org}
}

@inproceedings{peng2026correct,
  author       = {Yibo Peng and
                  James Song and
                  Lei Li and
                  Xinyu Yang and
                  Mihai Christodorescu and
                  Ravi Mangal and
                  Corina S. Pasareanu and
                  Haizhong Zheng and
                  Beidi Chen},
  editor       = {Maria Liakata and
                  Viviane P. Moreira and
                  Jiajun Zhang and
                  David Jurgens},
  title        = {When `Correct' Is Not Safe: Can We Trust Functionally Correct Patches
                  Generated by Code Agents?},
  booktitle    = {Proceedings of the 64th Annual Meeting of the Association for Computational
                  Linguistics (Volume 1: Long Papers), {ACL} 2026, San Diego, California,
                  United States, July 2-7, 2026},
  pages        = {15514--15546},
  publisher    = {Association for Computational Linguistics},
  year         = {2026},
  doi          = {10.18653/V1/2026.ACL-LONG.707},
  bibsource    = {dblp computer science bibliography, https://dblp.org}
}

@inproceedings{turpin2023language,
  author       = {Miles Turpin and
                  Julian Michael and
                  Ethan Perez and
                  Samuel R. Bowman},
  editor       = {Alice Oh and
                  Tristan Naumann and
                  Amir Globerson and
                  Kate Saenko and
                  Moritz Hardt and
                  Sergey Levine},
  title        = {Language Models Don't Always Say What They Think: Unfaithful Explanations
                  in Chain-of-Thought Prompting},
  booktitle    = {Advances in Neural Information Processing Systems 36: Annual Conference
                  on Neural Information Processing Systems 2023, NeurIPS 2023, New Orleans,
                  LA, USA, December 10 - 16, 2023},
  year         = {2023},
  bibsource    = {dblp computer science bibliography, https://dblp.org}
}

@article{chen2025reasoning,
  author       = {Yanda Chen and
                  Joe Benton and
                  Ansh Radhakrishnan and
                  Jonathan Uesato and
                  Carson Denison and
                  John Schulman and
                  Arushi Somani and
                  Peter Hase and
                  Misha Wagner and
                  Fabien Roger and
                  Vladimir Mikulik and
                  Samuel R. Bowman and
                  Jan Leike and
                  Jared Kaplan and
                  Ethan Perez},
  title        = {Reasoning Models Don't Always Say What They Think},
  journal      = {CoRR},
  volume       = {abs/2505.05410},
  year         = {2025},
  doi          = {10.48550/ARXIV.2505.05410},
  eprinttype   = {arXiv},
  eprint       = {2505.05410},
  bibsource    = {dblp computer science bibliography, https://dblp.org}
}

@inproceedings{sallou2024breaking,
  author       = {June Sallou and
                  Thomas Durieux and
                  Annibale Panichella},
  title        = {Breaking the Silence: the Threats of Using {LLMs} in Software Engineering},
  booktitle    = {Proceedings of the 2024 {ACM/IEEE} 44th International Conference on
                  Software Engineering: New Ideas and Emerging Results, NIER@ICSE 2024,
                  Lisbon, Portugal, April 14-20, 2024},
  pages        = {102--106},
  publisher    = {{ACM}},
  year         = {2024},
  doi          = {10.1145/3639476.3639764},
  bibsource    = {dblp computer science bibliography, https://dblp.org}
}

@article{ouyang2025empirical,
  author       = {Shuyin Ouyang and
                  Jie M. Zhang and
                  Mark Harman and
                  Meng Wang},
  title        = {An Empirical Study of the Non-Determinism of {ChatGPT} in Code Generation},
  journal      = {{ACM} Trans. Softw. Eng. Methodol.},
  volume       = {34},
  number       = {2},
  pages        = {42:1--42:28},
  year         = {2025},
  url          = {https://doi.org/10.1145/3697010},
  doi          = {10.1145/3697010},
  bibsource    = {dblp computer science bibliography, https://dblp.org}
}

@online{pillar2025rulesfile,
  author       = {Karliner, Ziv},
  title        = {New Vulnerability in {GitHub Copilot} and {Cursor}: How Hackers Can Weaponize Code Agents},
  organization = {Pillar Security},
  date         = {2025-03-18},
  url          = {https://www.pillar.security/blog/new-vulnerability-in-github-copilot-and-cursor-how-hackers-can-weaponize-code-agents},
  urldate      = {2026-05-21}
}

@online{layerx2026vibehacking,
  author       = {Paz, Roy},
  title        = {Vibe Hacking: {Claude Code} Can Be Turned Into a Nation-State-Level Attack Tool With No Coding at All},
  organization = {LayerX},
  date         = {2026-04-08},
  url          = {https://layerxsecurity.com/blog/vibe-hacking-claude-code-can-be-turned-into-a-nation-state-level-attack-tool-with-no-coding-at-all/},
  urldate      = {2026-05-21}
}

@online{nvidia2026agentsmd,
  author       = {Teixeira, Daniel},
  title        = {Mitigating Indirect {AGENTS.md} Injection Attacks in Agentic Environments},
  organization = {NVIDIA},
  date         = {2026-04-20},
  url          = {https://developer.nvidia.com/blog/mitigating-indirect-agents-md-injection-attacks-in-agentic-environments/},
  urldate      = {2026-05-21}
}

@online{anthropic2026memory,
    author={{Anthropic}},
    title={How {Claude} Remembers Your Project},
    date={2026-08-31},
    url={https://code.claude.com/docs/en/memory},
    urldate={2026-09-01}
}

@misc{openrouter,
    author={{OpenRouter}},
    title={{OpenRouter}},
    howpublished={\url{https://openrouter.ai}},
    note={Accessed: October 2026},
    year={2026},
}

@online{cwe_mitre,
  author  = {{MITRE}},
  title   = {{CWE} List Version 4.20},
  date    = {2026-04-30},
  version = {4.20},
  url     = {https://cwe.mitre.org/data/archive/},
  urldate = {2026-05-24}
}

@online{anthropic2026claudecode,
  author  = {{Anthropic}},
  title   = {{Claude Code}},
  date    = {2026-10-22},
  version = {2.1.308},
  url     = {https://github.com/anthropics/claude-code},
  urldate = {2026-10-22}
}

@online{anomaly2026opencode,
  author  = {{Anomaly}},
  title   = {{OpenCode}},
  date    = {2026-10-20},
  version = {2.1.1},
  url     = {https://github.com/anomalyco/opencode},
  urldate = {2026-10-22}
}

@online{anthropic2026sonnet55,
  author  = {{Anthropic}},
  title   = {Introducing {Claude Sonnet 5.5}},
  date    = {2026-09-28},
  url     = {https://www.anthropic.com/news/claude-sonnet-5-5},
  urldate = {2026-10-22}
}

@online{deepseek2026v41flash,
  author  = {{DeepSeek}},
  title   = {{DeepSeek-V4.1-Flash}: Smarter, Faster, More Efficient},
  date    = {2026-09-10},
  url     = {https://api-docs.deepseek.com/news/news260910/},
  urldate = {2026-10-22}
}

@online{qwen2026qwen38,
  author  = {{Qwen Team}},
  title   = {{Qwen3.8-27B}},
  date    = {2026-08-14},
  url     = {https://huggingface.co/Qwen/Qwen3.8-27B},
  urldate = {2026-10-22}
}

\end{document}